# Time and Frequency Injection into a Stabilized Fiber Link for Multi-clock Dissemination Network

WEI CHEN,[1] JIALIANG WANG,[2] QIN LIU,[2] NAN CHENG,[1] ZITONG FENG,[1] FEI YANG,[1] YOUZHEN GUI,[2,*] HAIWEN CAI[1,*]

[1]Shanghai Key Laboratory of All Solid-State Laser and Applied Techniques, Shanghai Institute of Optics and Fine Mechanics, Chinese Academy of Sciences, Shanghai, 201800, China
[2]Key Laboratory for Quantum Optics, CAS, Shanghai Institute of Optics and Fine Mechanics, Chinese Academy of Sciences, Shanghai, 201800, China
[3]Graduate University of the Chinese Academy of Sciences, Beijing 100049, China
*Corresponding author: yzgui@siom.ac.cn ; hwcai@siom.ac.cn

**Owing to the characteristics of ultra-low loss and anti-electromagnetic interference, using optical fiber to deliver time and frequency signal has been a preferred choice for high precise clock dissemination and comparison. As a brilliant idea, one has been able to reproduce ultra-stable signals from one local station to multiple users. In this paper, we take a step further. A concept of multi-clock (in different locations) dissemination for multi-terminals is presented. By injecting frequency signals into one stabilized ring-like fiber network, the relative stabilities of $3.4\times10^{-14}$@1 s for a master clock dissemination and $5.1\times10^{-14}$@1 s for a slave clock dissemination have been achieved. The proposed scheme can greatly simplify the future "N" to "N" time and frequency dissemination network, especially facing a multi-clock comparison situation.**

With the unprecedented progress of atomic clocks, the short term stability of a commercial cesium standard has achieved to $10^{-13}$ level, while the instability of an optical frequency reference is now better than $10^{-18}$[1-3]. To drastically utilize these ultra-stable reference, hundreds of clocks all over the world are linked and compared to each other to access their weighting factors, and that is the origin of International Atomic Time (TAI)[4]. The common methods of low loss coaxial cable for short distance linking and two-way satellite time and frequency transfer (TWSTFT) for long distance dissemination can no longer satisfy the performance that are currently achieved. Due to properties of low attenuation and anti-electromagnetic interference, optical fiber has been proved to be an efficient intermediary to delivery such high precise time and frequency signals by several research groups [5-8]. However, all the schemes are restricted in point to point transfer by round-trip method. Until recently, two solutions to get rid of the limit were proposed. One brilliant idea is about extracting signals anywhere along the trunk fiber [9-12]. The other effective implementation is to make the compensation structure serve in remote sites [13-15]. Both of the schemes bring great convenience for several users to access signals from one high precise clock. Whereas, a problem still leaves in all current schemes that one can only reproduce the ultra-stable signals from one local station to many users. In fact, a sound time and frequency distribution network should include a host of atomic clocks. That is to say, a more economic and easier "N" to "N" dissemination solution is still expected.

In this paper, we propose a scheme to address the problem, showing a strategy of injecting high precise time and frequency signals from different locations into a stabilized fiber network to meet the requirements of multi-clocks dissemination and comparison. With the solution, it can simplify the network structure where there are N stations including M clocks in different locations. As shown in Fig. 1(a), disseminating these M clock references (gray dots) to the N-M remote receivers (red dots) by the common point to point method, $M \cdot (N - M)$ segments of fiber are needed in the network. Improved a lot, if the scheme of multi-access technique is chosen, like Fig. 1(b), the cost can be reduced to M, but it still requires M independent active noise compensation servers. Here, in Fig. 1 (c), with the proposed scheme all stations can be connected to each other in one ring-like fiber while only one active noise compensator is needed. Namely, each recovery sites can reproduce the time and frequency signals from different clocks at the same time with the ring-like structure and special noise suppression strategy.

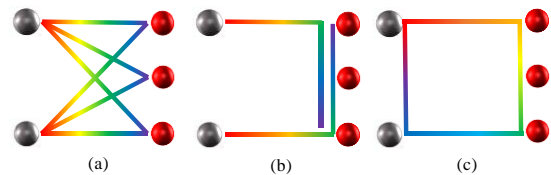

Fig.1 Example of dissemination methods in multipoint network showing the cost of fiber links: (a), point to point method; (b), multi-access technique; (c), proposed ring-like injection solution

To explain the basic ideas of ring-like injection, we first introduce the structure of one master clock station with one recovery terminal. That is a particular case of the scheme where N=2, and M=1, seems to be a point to point condition. However different from the round-trip method [16-17], as shown in Fig. 2, we bi-directionally distribute the ultra-stable frequency signal into a fiber optic link in the master clock station. The fiber optic link is shaped into a ring-like structure. This structure can make it easier for other clocks to disseminate the stable reference which we will discuss later. In realistic world, it can be two parallel fiber that makes the light transmit in both directions.

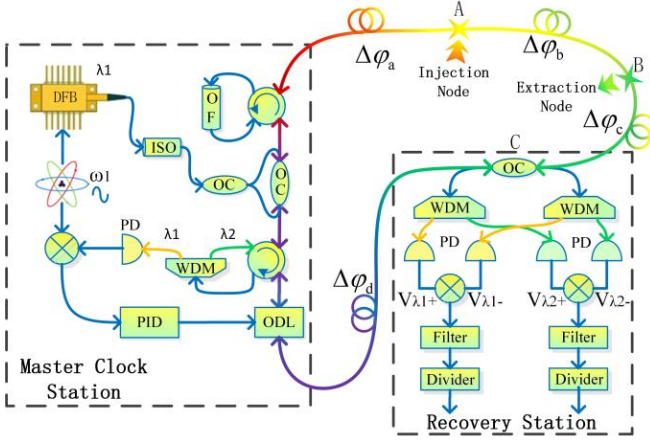

Fig. 2. Outlines of the multi-clocks dissemination system. DFB: distributed feedback laser, ISO: isolator, WDM: wavelength division multiplexer, OC: optical coupler, OF: optical filter, ODL: optical delay line, PID: proportional-integral-derivative, PD: photodetector

Focusing on the master clock station, to jointly suppress the phase noise for all clocks references, optical compensation method is chosen to be the suppression actuator. Here we use an optical delay line [14, 17] including a temperature controlled ring and a fast fiber stretcher to stabilize the whole optical link. The master clock frequency reference modulates a distributed feedback (DFB) laser with wavelength of $\lambda_1$. Without considering its amplitude, it can be expressed as:

$$V_{\lambda 1} = \cos(\omega_1 t + \varphi_1) \quad (1)$$

The ring-like fiber is artificially divided into four sections. Clockwise signal passes through four sections of the fiber link, successively suffering from phase noise of $\Delta\varphi_a$, $\Delta\varphi_b$, $\Delta\varphi_c$ and $\Delta\varphi_d$. When it returns back to master clock station, an error signal can be obtained after a phase discriminator:

$$\Delta\varphi_{err} = \Delta\varphi_a + \Delta\varphi_b + \Delta\varphi_c + \Delta\varphi_d \quad (2)$$

Through a proportional-integral-derivative (PID) process, we apply this error signal to drive the optical delay line, making it actuate a phase correction of $\Delta\varphi_{ODL} = -\Delta\varphi_{err}$. That means the clockwise signal is stabilized after a full circle.

$$\Delta\varphi_a + \Delta\varphi_b + \Delta\varphi_c + \Delta\varphi_d + \Delta\varphi_{ODL} = 0 \quad (3)$$

In the meantime, the anticlockwise signal propagates in the same fiber link but in reverse direction. While returning back to the master station, an optical filter (OF) is used to avoid the propagating light of $\lambda_1$ back into the fiber link repeatedly. Suffering the common noise fluctuation terms, the total phase drifting of the anticlockwise signal over one round is also eliminated.

The recovery station can be inserted anywhere along the fiber link. Here we take point C which is 25km away from the master clock as illustration. Shown in Fig. 2, an optical couple (OC) is used to download the light in two directions from the circle link. Dense wavelength division multiplexing technology is applied to extract the light of $\lambda_1$ from the master clock station (yellow line). Other wavelength channels are left for slave clocks. The clockwise and anticlockwise signal extracted in this station can be expressed respectively as:

$$V_{\lambda 1+} = \cos(\omega_1 t + \varphi_1 + \Delta\varphi_a + \Delta\varphi_b + \Delta\varphi_c) \quad (4)$$

$$V_{\lambda 1-} = \cos(\omega_1 t + \varphi_1 + \Delta\varphi_{ODL} + \Delta\varphi_d) \quad (5)$$

They are then mixed by a balanced mixer and filtered to remove the direct current out. An up-converted frequency signal is gained:

$$V_{2\omega 1} = \cos(2\omega_1 t + 2\varphi_1 + \Delta\varphi_a + \Delta\varphi_b + \Delta\varphi_c + \Delta\varphi_d + \Delta\varphi_{ODL}) \quad (6)$$

According to equation 3, all of the phase fluctuation terms are cancelled out. In order to keep the same radio frequency with the master clock, a divider is used to achieve a stable frequency:

$$V_{\omega 1} = \cos(\omega_1 t + \varphi_1 + \xi_1) \quad (7)$$

where $\xi_1$ is the intrinsic phase shift induced by the frequency divider. Thus the frequency reference of the master clock is duplicated in the recovery station.

To expand the application, the key advantage of the scheme is that it allows to inject other clock references into the stabilized fiber optic link. The structure of a slave clock injection station is shown in Fig. 3. Theoretically, it can also be placed at arbitrary node along the ring-like network. As an experimental test, a 2km fiber spool is adopted between the master station and the slave station (at point A in Fig. 2.).We use another wavelength of $\lambda_2$ to carry the frequency of the slave clock. The original signal can be written as:

$$V_{\lambda 2} = \cos(\omega_2 t + \varphi_2) \quad (8)$$

Similar to the master clock station, the light carrying the ultra-stable frequency is transmitted into the fiber link in two directions. The signals pass through one round of the fiber link and then come back to the slave station. They will no longer enter the link once again because the two optical filter are used to stop the wavelength of $\lambda_2$ passing through. This design prevents the light superimposing on the propagating signal, otherwise it may seriously degrade the signal-to-noise ratio upon photodetector.

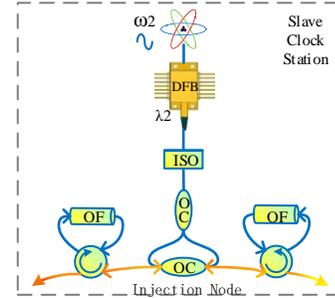

Fig.3. Schematic diagram of the slave clock injection station. OC: optical coupler, OF: optical filter

Noting that thanks to the ring-like injection structure as we proposed, the light of $\lambda_2$ passes through the common segments of the fiber link, suffering almost the same impact of vibration and temperature changing with light of $\lambda_1$. So an optical noise suppression actuator can jointly compensate the phase noise imposed on them. That is to say the optical delay line in master clock station is still effective for the slave clock signal. Here ignoring noise within a clock station, the total phase drifting of the slave clock reference after one circle propagation is equal to that of equation 3.

In the recovery station, the other wavelength channel of WDM (green line) is coming in handy. The slave clock frequency signal from two directions are:

$$V_{\lambda 2+} = \cos(\omega_2 t + \varphi_2 + \Delta\varphi_b + \Delta\varphi_c) \quad (9)$$

$$V_{\lambda 2-}=\cos(\omega_2 t+\varphi_2+\Delta\varphi_a+\Delta\varphi_{ODL}+\Delta\varphi_d) \quad (10)$$

Even though each of them is unstable and their phase fluctuation term are different from $V_{\lambda 1+}$ and $V_{\lambda 1-}$, the mixed signal $V_{2\omega 2}$ is analogous to equation 6.

$$V_{2\omega 2}=\cos(2\omega_2 t+2\varphi_2+\Delta\varphi_a+\Delta\varphi_b+\Delta\varphi_c+\Delta\varphi_d+\Delta\varphi_{ODL}) \quad (11)$$

In fact, the phase drifting terms also sum to be zero. After the frequency divider, the recovered output frequency is stabilized and obtained.

$$V_{\omega 2}=\cos(\omega_2 t+\varphi_2+\xi_2) \quad (12)$$

where $\xi_2$ is the corresponding intrinsic phase shift induced by the divider. Finally all the recovered frequency references thus click the same as they are in the clock stations.

To demonstrate the idea of "N" to "N" dissemination scheme, we insert another identical recovery station in point B which is 20km away from point A and 5km away from point C. Like some of our previous work [14], the fibers that contribute to phase fluctuations in each stations are set as short as we can and put into a case that full of sponges, aiming to reduce the residual phase noise induced by them. The relative frequency stabilities of both recovered signals are measured by comparing them with their original signals, such as we compare $V_{\omega 2}$ to $V_{\lambda 2}$ to evaluate the performance of slave clock dissemination. Both recovery stations play a very similar behavior of the overlapping Allan deviation (ADEV). To be clear, here we only demonstrate the stability curves of the recovery station at point C as an illustration. The results are shown in Fig. 4. Short-time instability can reach $3.4\times10^{-14}$@1 s for the master clock dissemination and $5.1\times10^{-14}$@1 s for the slave clock dissemination, while long-time performance extends to $5.5\times10^{-17}$ @$10^4$ s and $8.6\times10^{-17}$ @$10^4$ s respectively.

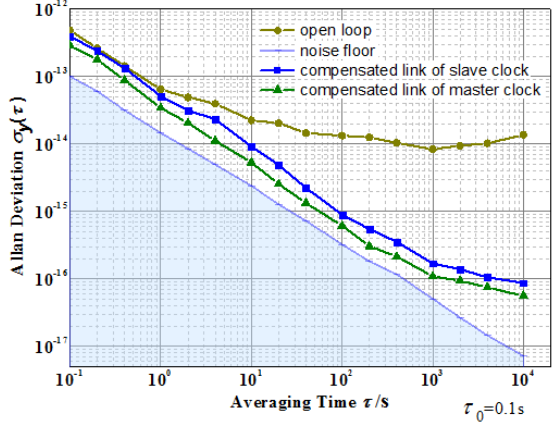

Fig.4. Relative stabilities of frequency references for different clock stations

Compared with free-running state, both of them improve a lot, especially in long term averaging time. In general, the trends of two curves are similar to each other but the slave clock one is slightly worse for the whole time scale. That is reasonable because of the two mainly causes: a) Even though the optical path for the two modulated lights are almost the same, they still have some uncompensated parts, such as the fiber in each stations and the noise induced by electro-optic conversion process. These parts of noise are out of the feedback control loop, which will introduce additional noise. b) As long as the dispersion exists, the error signal which is used to drive the optical delay line corresponding to $\lambda_1$ cannot be completely equal to that of the light of $\lambda_2$. Indeed, the thermal sensitivity of the phase-delay deviation with respect to dispersion is different. A simple solution can be adding a dispersion compensation fiber of adequate length into the ring-like link. Apart from the holistic behavior between two curves, the slope turns to be slow after 100s. It is principally because the mixers we used are nonlinearity and sensitive to temperature changing even the recovery terminal has been placed into a case. Nevertheless, the performance of the compensated state proves the scheme to be effective

As for network design, another advantage of these scheme is that a recovery terminal can be set in the same lab of master clock station, since the recovery station is not sensitive to locations. It is helpful for an administrator to monitor all the performance of clocks existed in the network. Through this monitor we can realize which atomic clock run best. Then, users do not have to recover all the clock references. As the performance are known, it is convenient for users to choose the best one to be recovered by using a wavelength selective switch (WSS) instead of the WDM in the recovery stations.

In some particular application, such as clock comparison, we only need to obtain the frequency (or phase) difference between different clocks. The proposed scheme can be also effective, even though the optical fiber path is time-varying. For further illustration, if these two clocks are in one laboratory, the phase of equation 1 can be directly compared with the phase of equation 8:

$$\Delta\varphi=(\omega_1 t-\omega_2 t)+(\varphi_1-\varphi_2) \quad (13)$$

However, in reality, high precise atomic clocks run in different institutions. We can apply the clock injection structure as we proposed above, but turn off the noise suppression servo in the master clock station. At the recovery station the output frequency of each clocks should be modified to

$$V'_{\omega 1}=\cos(\omega_1 t+\varphi_1+\frac{\Delta\varphi_a+\Delta\varphi_b+\Delta\varphi_c+\Delta\varphi_d+\Delta\varphi_{ODL}}{2}) \quad (14)$$

$$V'_{\omega 2}=\cos(\omega_2 t+\varphi_2+\frac{\Delta\varphi_a+\Delta\varphi_b+\Delta\varphi_c+\Delta\varphi_d+\Delta\varphi_{ODL}}{2}) \quad (15)$$

Even if the phase fluctuation term cannot sum to zero this time, the result of comparing the phase difference of these two signals is the same as equation 13. If there are only two clocks, it is similar to TWTFT strategy [18-19]. But it can be exciting when facing a multi-clocks situation. All the clocks can be compared with each other at the same time without considering the common phase noise imposed by the fiber as if they are in one laboratory.

In conclusion, we propose a novel "N" to "N" time and frequency dissemination scheme. It is fantastic that frequency references from different clock stations are allowed to inject anywhere along a ring-like fiber link and can be jointly recovered without using any extra noise suppression actuator. It dramatically simplifies the complexity of a multi-clocks dissemination network. The ability of network management can be enhanced by monitoring the performance of all the atomic clocks hanging upon the link. As a proof-of-principle experimental test, two clock signals are injected into a stabilized fiber optic link up to 52km. The instabilities of $3.4\times10^{-14}$@1 s and $5.5\times10^{-17}$ @$10^4$ s for the master clock and $5.1\times10^{-14}$@1 s and $8.6\times10^{-17}$ @$10^4$ s for the slave clock are obtained. Like the common round-trip method, the proposed scheme is not restricted to transfer radio frequency signal. The basic idea can be extended to optical frequency delivery and pulse transmission. In addition, we introduce a special application for multi-clock comparison via a time-varying fiber path. We expect this scheme is applicable for the time and frequency synchronization network which are currently under construction.

**Acknowledgment**. We would like to thank Professor Zujie Fang for helpful discussion and guidance.